\documentclass[a4paper,twocolumn,11pt,unpublished]{quantumarticle}
\pdfoutput=1
\usepackage[utf8]{inputenc}
\usepackage[english]{babel}
\usepackage[T1]{fontenc}

\usepackage{amsmath,amssymb,amsthm}
\usepackage{graphicx}
\usepackage{booktabs}
\usepackage{microtype}
\usepackage{enumitem}
\usepackage{float}
\usepackage{subcaption}
\usepackage{tcolorbox}
\usepackage{mathtools}
\usepackage{multirow}
\usepackage[numbers,sort&compress]{natbib}
\usepackage{hyperref}

\definecolor{boxbg}{RGB}{240,248,255}
\definecolor{boxrule}{RGB}{70,130,180}
\definecolor{notegreen}{RGB}{0,100,0}

\tcbuselibrary{skins,breakable}

\newtcolorbox{keybox}[1][]{%
  enhanced, breakable,
  colback=boxbg, colframe=boxrule,
  fonttitle=\bfseries, title=#1,
  left=6pt, right=6pt, top=4pt, bottom=4pt
}

\newtcolorbox{notebox}{%
  enhanced, breakable,
  colback=yellow!10, colframe=orange!60,
  left=6pt, right=6pt, top=4pt, bottom=4pt
}

\newtheorem{theorem}{Theorem}[section]
\newtheorem{proposition}[theorem]{Proposition}

\newtheorem{remark}[theorem]{Remark}

\newcommand{\R}{\mathbb{R}}
\newcommand{\C}{\mathbb{C}}
\newcommand{\norm}[1]{\left\|#1\right\|}
\newcommand{\abs}[1]{\left|#1\right|}
\newcommand{\bA}{\mathbf{A}}
\newcommand{\bb}{\mathbf{b}}
\newcommand{\bu}{\mathbf{u}}
\newcommand{\bv}{\mathbf{v}}
\newcommand{\lmin}{\lambda_{\min}}
\newcommand{\lmax}{\lambda_{\max}}
\newcommand{\kap}{\kappa}

\begin{document}
\title{Spectrally Corrected Polynomial Approximation
for Quantum Singular Value Transformation}

\author{Krishnan Suresh}
\email{ksuresh@wisc.edu}
\orcid{0000-0002-9688-9697}
\affiliation{University of Wisconsin--Madison, Madison, WI 53706, USA}

\maketitle
\begin{abstract}
Quantum Singular Value Transformation (QSVT) provides a unified framework for applying polynomial functions to the singular values of a block-encoded matrix. QSVT prepares a state
proportional to $\bA^{-1}\bb$ with circuit depth
$O(d\cdot\mathrm{polylog}(N))$, where $d$ is the polynomial
degree of the $1/x$ approximation and $N$ is the size of $\bA$. Current polynomial approximation  methods are over the continuous interval $[a,1]$, giving $d = O(\sqrt{\kap}\log(1/\varepsilon))$, and make no use of any
properties of $\bA$.

We observe here that QSVT solution accuracy depends only on the
polynomial accuracy at the eigenvalues of $\bA$.
When all $N$ eigenvalues are known exactly, a pure spectral polynomial $p_{S}$ 
can interpolate $1/x$ at these eigenvalues and achieve unit fidelity at reduced degree. But its practical applicability is limited. To address this, we propose a spectral
correction that exploits prior knowledge of $K$ eigenvalues of
$\bA$. Given any base polynomial $p_0$, such as Remez, of degree $d_0$, a
$K\times K$ linear system enforces exact interpolation of
$1/x$ only at these $K$ eigenvalues without increasing
$d_0$. The spectrally corrected polynomial $p_{SC}$ preserves the continuous error profile
between eigenvalues and inherits the parity of $p_0$.

QSVT experiments on the 1D Poisson equation demonstrate up to a
$5\times$ reduction in circuit depth relative to the base
polynomial, at unit fidelity and improved compliance error. The
correction is agnostic to the choice of base polynomial and robust
to eigenvalue perturbations up to $10\%$ relative error. Extension
to the 2D Poisson equation suggests that correcting a small
fraction of the spectrum may suffice to achieve fidelity above
$0.999$.
\end{abstract}

\section{Introduction}
\label{sec:intro}

Quantum Singular Value Transformation (QSVT), introduced by Gily\'{e}n
et al.~\cite{gilyen2019quantum}, provides a unified framework for applying
polynomial functions to a matrix encoded as a quantum oracle \cite{martyn2021grand}. Its most
celebrated application is the quantum linear systems
algorithm~\cite{childs2017quantum}: given a block encoding
of a Hermitian positive definite (HPD) matrix $\bA$ of size $N \times N$, and a polynomial $p$ of degree $d$ approximating $1/x$, QSVT prepares a quantum state proportional to
$\bA^{-1}\bb$ with circuit depth $O(d\cdot\mathrm{polylog}(N))$. All existing polynomial construction methods --- the Remez minimax polynomial~\cite{trefethen2019approximation}, the Mang truncated
Chebyshev series~\cite{mang2024}, and analytical construction by 
S\"{u}nderhauf et al.~\cite{sunderhauf2025matrix} --- treat the approximation
over the \emph{continuous} interval $[a,1]$, where $a = \lmin(\bA)$, giving
degree $O(\sqrt{\kap}\log(1/\varepsilon))$ with $\kap = \lmax/\lmin$ and
target accuracy $\varepsilon$. They make no use of any properties of
$\bA$. 

In this paper, we investigate whether the \emph{spectral}
structure of $\bA$ can be exploited to improve the polynomial
approximation. We show that if $K \ll N$ eigenvalues of $\bA$ are
known, either analytically or approximately, a simple correction
to any base polynomial yields significant improvements in QSVT
solution accuracy at no increase in circuit depth. Specifically, we derive a 
simple min-norm correction to any base polynomial by enforcing exact
interpolation at $K$ specified eigenvalues while minimizing the perturbation
to the Chebyshev coefficients. The correction only requires a $K\times K$ linear solve.

Section~\ref{sec:qsvt} reviews block encoding in QSVT, the polynomial approximation problem, and the three aforementioned base polynomials. Section~\ref{sec:spectralpoly} introduces the pure \emph{spectral polynomial} where we disregard the base polynomials and construct a polynomial that exactly interpolates all the $N$ eigenvalues as a minimum-$L^2$-norm solution to an under-determined system. This, however, suffers from several deficiencies. Section~\ref{sec:spectralCorrected} presents a more practical \emph{spectrally corrected} method that provides eigenvalue corrections to a base polynomial, leading to spectral-Remez, spectral-Mang, and spectral-S\"{u}nderhauf polynomials. Section~\ref{sec:experiments} presents several numerical experiments involving these spectrally corrected polynomials in QSVT. Section~\ref{sec:discussion} discusses the scope, limitations, and future directions of this work. This is followed by conclusions in Section~\ref{sec:conclusion}.

\section{Technical Background}
\label{sec:qsvt}

\subsection{Block encoding}
\label{sec:block_enc}

A \emph{block encoding} of $\bA \in \R^{N \times N}$ is a unitary
$U_{\bA} \in \C^{2^a N \times 2^a N}$ such that~\cite{gilyen2019quantum}
\[
  \bA / \alpha = (\langle 0^a| \otimes I)\, U_{\bA}\, (|0^a\rangle \otimes I),
\]
where $a$ is an ancilla qubit count and $\alpha \geq \norm{\bA}$ is a
sub-normalization factor, and $\bA/\alpha$ appears in the top-left 
block of $U_{\bA}$:
\begin{align*}
    U_{\bA} =
    \begin{bmatrix}
    \bA / \alpha & \cdot \\
    \cdot & \cdot
    \end{bmatrix}
\end{align*}
For sparse matrices with at most $s$ non-zeros per row,
a block encoding can be constructed using $O(s\,\mathrm{polylog}(N))$
elementary gates~\cite{childs2017quantum}. 

\subsection{Polynomial approximation}
\label{sec:polyapprox}

Besides block encoding, the second critical ingredient in QSVT is a polynomial approximation $p(x)$
to $1/x$ over the interval $[a, 1]$, where $a = \lmin$, and $\lmin$ is lowest singular value of $\bA/\alpha$. Three requirements are imposed on polynomial approximations:
\begin{enumerate}
    \item \textbf{Parity}: $p(x)$ must be an odd polynomial, i.e.\ $p(-x) = -p(x)$.
    \item \textbf{Boundedness}: $|p(x)| \leq 1$ for all $x \in [-1,1]$.
    In practice, a polynomial $\hat{p}(x)$ is first constructed to
    approximate $1/x$ and then normalized:
    $p(x) = \hat{p}(x)/\tau$, where
    \begin{equation}
      \tau = \max_{x \in [-1,-a] \cup [a,1]} |\hat{p}(x)|
      \label{eq:tauDefinition}
    \end{equation}
    is the \emph{polynomial subnormalization factor}. The
    post-selection success probability of QSVT scales as
    \begin{equation}
      P_{\rm succ} = \frac{\|p(\bA)\bb\|^2}{\tau^2},
      \label{eq:psucc}
    \end{equation}
    so a large $\tau$ reduces the probability of obtaining the
    desired output state upon measurement.
    \item \textbf{Accuracy}: $p(x)$ should approximate $1/x$ to a
    desired tolerance $\varepsilon$:
\begin{equation}
  \abs{x\,p(x) - 1} \leq \varepsilon
  \quad \forall\, x \in [a, 1].
  \label{eq:approx_goal}
\end{equation}
\end{enumerate}

Here, we consider three methods for constructing such polynomials:  Remez~\cite{trefethen2019approximation}, Mang~\cite{mang2024}, and S\"{u}nderhauf~\cite{sunderhauf2025matrix}. 
All three methods represent $p(x)$ in the odd Chebyshev basis:
\begin{align}
\label{eq:chebApproximation}
  p(x) = \sum_{j=0}^{\lfloor d/2 \rfloor} c_j\, T_{2j+1}(x),
\end{align}
where $T_k(x) = \cos(k\,\arccos x)$ are Chebyshev polynomials  of the first
kind \cite{trefethen2019approximation}.  All three methods satisfy the parity requirement by construction. However, they differ in the minimum degree $d$ needed to achieve the desired  $\varepsilon$ accuracy, and how the coefficients $\{c_j\}$ are computed. Remez 
enforces accuracy in the $L^\infty$ sense over $[a,1]$; Mang minimizes an $L^2$ residual in the $\theta$-parametrization, which satisfies 
\eqref{eq:approx_goal} in practice, but without a uniform guarantee; and S\"{u}nderhauf provides a closed-form analytical certificate for both boundedness and accuracy simultaneously.

\subsubsection{Remez polynomial}
\label{sec:remez}

The Remez algorithm~\cite{trefethen2019approximation}, proposed by Remez (1934), finds the polynomial minimizing the
maximum error over the continuous interval:
\begin{equation}
  p^* = \arg\min_{p \in \mathcal{P}_d^{\rm odd}}
        \max_{x \in [a,1]} \abs{x\,p(x) - 1}.
  \label{eq:minimax}
\end{equation}
By the Chebyshev equioscillation theorem \cite{trefethen2019approximation}, the solution achieves the same
maximum error at $d+2$ points in $[a,1]$ with alternating signs. The algorithm requires solving increasingly ill-conditioned Vandermonde systems as $d$ grows. Since $d$ scales as $O(\sqrt{\kap}\log(1/\varepsilon))$, high-$\kap$ problems may require extended-precision arithmetic.   The equioscillatory property, while optimal for continuous approximation, leaves \emph{no slack}: the error touches $\varepsilon$
at $d+2$ points with alternating signs, so any perturbation to the coefficients immediately violates the uniform bound at those points. This will be of relevance in Section \ref{sec:spectralCorrected}.

\subsubsection{Mang polynomial}
\label{sec:mang}

Mang et al.~\cite{mang2024} re-parametrize the problem via $x = \cos\theta$ and minimize the
$L^2$ residual over $\theta \in [0, \arccos a]$:
\begin{equation}
  \min_{\mathbf{c}} \int_0^{\arccos a}
  \left| \cos\theta \sum_{j=0}^{n-1} c_j\, T_{2j+1}(\cos\theta) - 1 \right|^2
  d\theta,
  \label{eq:mang_ls}
\end{equation}
which reduces to a linear least-squares problem for the Chebyshev
coefficients $\mathbf{c} \in \R^n$, discretized on a uniform $\theta$-grid.
The degree $d = 2n-1$ is chosen so that the resulting polynomial
satisfies~\eqref{eq:approx_goal}. The $L^2$ objective does not guarantee uniform
accuracy: the error $|xp(x)-1|$ is not equioscillatory and can peak near
$x = a$. The degree scales as $O(\sqrt{\kap}\log(1/\varepsilon))$ and is
not guaranteed to be lower than Remez; the relative degree depends on
the specific values of $\kap$ and $\varepsilon$.

\subsubsection{S\"{u}nderhauf polynomial}
\label{sec:sunderhauf}

S\"{u}nderhauf et al.~\cite{sunderhauf2025matrix} derive a \emph{closed-form}
odd polynomial that natively satisfies the boundedness requirement
$|p(x)| \leq 1$ on $[-1,1]$, without any iterative
optimization or linear solve. The polynomial is expressed directly in terms of the
problem parameters $a$ and $\varepsilon$, making it the only method among the three that analytically satisfies all three requirements. Furthermore, the method is numerically stable at a high degree, but the degree is often higher than Remez.

\subsubsection{Comparison}
\label{sec:methods_summary}

\begin{figure}[H]
  \centering
  \subcaptionbox{Approximating functions.\label{fig:poly_approx}}
    {\includegraphics[width=1\linewidth]{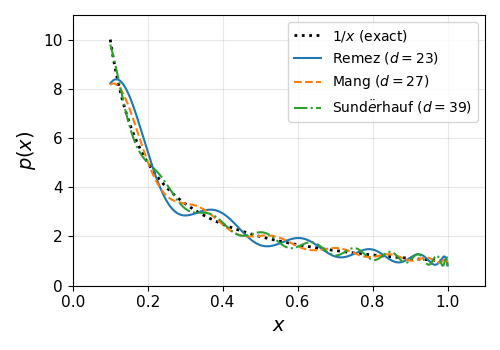}}
  \hfill
  \subcaptionbox{Pointwise error.\label{fig:poly_error}}
    {\includegraphics[width=1\linewidth]{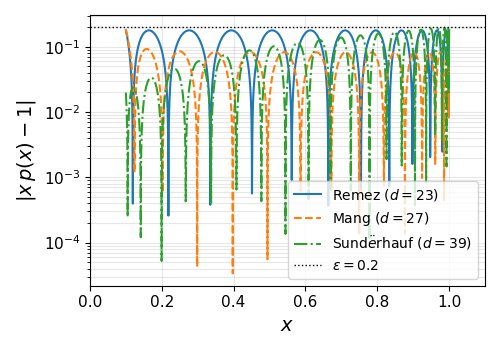}}
  \caption{Polynomial approximation to $1/x$ for $\kappa=10$,
  $\varepsilon=0.2$. (a) Approximation $p(x)$ vs $1/x$. (b) Point-wise
  error $|xp(x)-1|$: Remez ($d=23$) achieves a uniform equioscillatory
  error; Mang ($d=27$) peaks near $x=a$; S\"{u}nderhauf ($d=39$)
  satisfies the bound analytically. A large $\varepsilon$ is used for
  visual clarity; in practice $\varepsilon\sim0.01$.}
  \label{fig:polynomialComparison}
\end{figure}

Figure \ref{fig:polynomialComparison}a provides a visual comparison of the three approximations for $\kappa = 10$ and $\varepsilon = 0.2$. A large $\varepsilon$ is used for visual clarity; in practice $\varepsilon\sim0.01$. Remez requires the lowest degree ($d = 23$), followed by Mang ($d = 27$) and then S\"{u}nderhauf ($d = 39$). In practice, for typical values of $\kappa = 100$ and $\varepsilon = 0.01$, $d$ can exceed 1000.
  
 Figure~\ref{fig:polynomialComparison}b confirms the theoretical properties of
each method. Remez exhibits the classical equioscillation predicted
by Chebyshev's theorem~\cite{trefethen2019approximation}: the error touches
$\varepsilon$ at equal height across $[a,1]$, achieving the minimum possible
degree for a given $\varepsilon$. Mang shows a similar oscillatory
pattern but with peaks slightly higher near $x=a$, consistent with the
$L^2$ objective under-weighting the left endpoint. S\"{u}nderhauf 
stays strictly below $\varepsilon$ everywhere, confirming it is a
conservative analytical bound, with a higher degree ($d=39$ vs $d=23$ for
Remez).

\section{Pure Spectral Polynomial}
\label{sec:spectralpoly}

The three methods reviewed in Section~\ref{sec:polyapprox} share a
common philosophy: construct a polynomial that approximates $1/x$
over the entire continuous interval $[a,1]$. From a QSVT
perspective, however, this is stronger than necessary. The QSVT
output state is proportional to
$p(\bA)\bb = \sum_k p(\lambda_k)(\bv_k^T\bb)\bv_k$, and the
error relative to
$\bA^{-1}\bb = \sum_k \lambda_k^{-1}(\bv_k^T\bb)\bv_k$ depends
only on the values $\{p(\lambda_k)\}$, not on $p(x)$ between
eigenvalues. The necessary and sufficient accuracy condition is
discrete:
\begin{equation}
  \max_{k=1,\ldots,N} \abs{\lambda_k\,p(\lambda_k) - 1}
    \leq \varepsilon.
  \label{eq:discrete_goal}
\end{equation}

If all $N$ eigenvalues of $\bA$ are known, the continuous
approximation problem reduces to a finite interpolation problem.
Let $p$ be represented in the odd Chebyshev basis with $n$ terms:
\[
  p(x) = \sum_{j=0}^{n-1} c_j\, T_{2j+1}(x).
\]
The $N$ interpolation constraints
$\lambda_k\,p(\lambda_k) = 1$ form the linear system
\begin{equation}
  \underbrace{(\Lambda \mathbf{B})}_{N \times n} \mathbf{c}
    = \mathbf{1},
  \label{eq:interpolation_system}
\end{equation}
where
$B_{kj} = T_{2j+1}(\lambda_k)$ and
$\Lambda = \mathrm{diag}(\lambda_1,\ldots,\lambda_N)$.
When $n > N$ the system is underdetermined; we select the
minimum-$\ell^2$-norm solution via the Moore--Penrose
pseudoinverse~\cite{golub2013matrix}:
\begin{equation}
  \mathbf{c}^* = (\Lambda\mathbf{B})^+ \mathbf{1}.
  \label{eq:minnorm}
\end{equation}
The resulting \emph{spectral polynomial} has degree
$d = 2n - 1$ with
$n = \lceil n_{\rm factor} \cdot N \rceil$ for a user-chosen
ratio $n_{\rm factor} \geq 1$. The minimum-norm solution
selects the smoothest polynomial that exactly interpolates
$1/\lambda_k$ at all $N$ eigenvalues.

\begin{figure}[htbp]
  \centering
  \includegraphics[width=0.9\columnwidth]{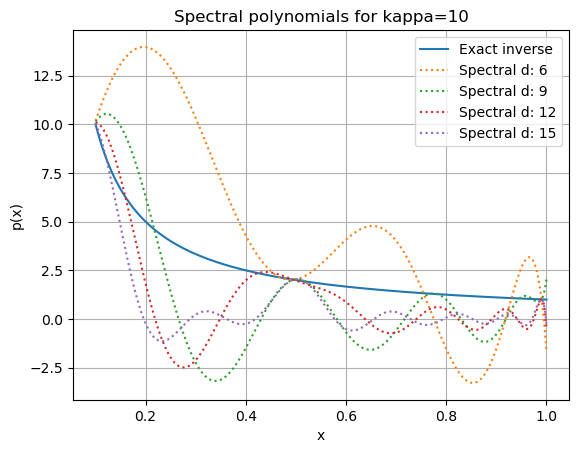}
  \caption{Pure spectral polynomial $p_S(x)$ for $\kappa=10$
  with $N=3$ eigenvalues at $\lambda = 0.1, 0.5, 1.0$, at
  increasing degree. All polynomials interpolate $1/\lambda_k$
  exactly at the three eigenvalues. As the degree increases, the
  inter-eigenvalue oscillations diminish and $\tau$ approaches
  $\kappa$.}
  \label{fig:spectral_poly}
\end{figure}

Figure~\ref{fig:spectral_poly} illustrates the spectral polynomial
for $\kappa = 10$ with three eigenvalues at
$\lambda = 0.1, 0.5, 1.0$. At the minimum degree ($d = 5$,
$n_{\rm factor} = 1$), the polynomial interpolates
$1/\lambda_k$ exactly but oscillates between eigenvalues,
producing a large $\tau$ (see \ref{eq:tauDefinition}). As the degree increases, the additional
degrees of freedom allow the minimum-norm solution to suppress
these oscillations. More importantly, since only the values at the eigenvalues
affect the QSVT output, the oscillations do not degrade fidelity
--- but they inflate $\tau$ and reduce the success probability
$P_{\rm succ} \propto 1/\tau^2$.

While the spectral polynomials exactly interpolate the eigenvalues (and therefore  result in zero QSVT error), they suffer from three practical limitations:
\begin{enumerate}
\item \textbf{All eigenvalues required.}
  The polynomial provides no guarantee at uncorrected
  eigenvalues. If only $K < N$ eigenvalues are known, the
  uncontrolled oscillations at uncorrected eigenvalues can
  destroy solution accuracy.
\item \textbf{Degree grows with $N$.}
  The degree $d = 2\lceil n_{\rm factor}\cdot N\rceil - 1$
  grows linearly in $N$, and the ratio $n_{\rm factor}$ required
  for $\tau \approx \kappa$ increases with $\kappa$, eventually
  making the degree comparable to the base polynomial degree
  $d = O(\sqrt{\kappa}\log(1/\varepsilon))$.
\item \textbf{Sensitivity to eigenvalue errors.}
  Without a continuous approximation guarantee, any discrepancy
  between the assumed and true eigenvalues produces errors that
  are unbounded by $\varepsilon$. In contrast, a base polynomial
  ensures $|\lambda\,p(\lambda) - 1| \leq \varepsilon$ for all
  $\lambda \in [a,1]$, regardless of eigenvalue accuracy.
\end{enumerate}
These limitations motivate a correction strategy discussed next: use a base
polynomial to provide a continuous safety net, and correct it at
the $K$ eigenvalues that matter most.

\section{Spectrally Corrected Polynomial}
\label{sec:spectralCorrected}
\subsection{The min-norm correction}
\label{sec:correction}
A more practical strategy starts from any base polynomial $p_0$
(Remez, Mang or S\"{u}nderhauf) and creates a spectrally
corrected polynomial $p_{SC} = p_0 + p_{\rm corr}$ that satisfies
$\lambda_k\, p_{SC}(\lambda_k) = 1$ at the $K$ smallest eigenvalues
$\lambda_1 \leq \cdots \leq \lambda_K$. 

\begin{remark}
Although the proposed correction applies to any subset of eigenvalues, the
contiguous $K$ smallest are preferred for two reasons: (1) if an uncorrected eigenvalue lies between a pair of targeted eigenvalues, the corrected polynomial could exhibit a large error at the uncorrected eigenvalue,  and (2) the lowest eigenvalues are often the most critical in many engineering settings. 
\end{remark}

The correction polynomial $p_{\rm corr}$ is represented in the same odd Chebyshev basis as
$p_0$:
\begin{align*}
  p_{\rm corr}(x) = \sum_{j=0}^{n_0-1} \Delta c_j\, T_{2j+1}(x),
\end{align*}
where $\Delta \mathbf{c} \in \R^{n_0}$ is the perturbation to the
Chebyshev coefficients. By enforcing the $K$ constraints, the coefficients are determined as a min-norm correction.

\begin{keybox}[Spectral correction]
\textbf{Step 1.}  Compute the residuals at the $K$ smallest
eigenvalues:
\begin{align*}
  r_k = 1 - \lambda_k\,p_0(\lambda_k), \qquad k = 1,\ldots,K.
\end{align*}
\textbf{Step 2.}  Solve the $K \times K$ Gram system for
multipliers $\boldsymbol\alpha \in \R^K$:
\begin{align}
  &\mathbf{G}\,\boldsymbol\alpha = \mathbf{r},
  \label{eq:gram}\\
  &G_{ij} = \lambda_i\lambda_j
    \sum_{\ell=0}^{n_0-1}
    T_{2\ell+1}(\lambda_i)\,T_{2\ell+1}(\lambda_j),
  \nonumber
\end{align}
where $n_0 = (d_0 + 1)/2$ is the number of odd Chebyshev terms
in the base polynomial $p_0$ of degree $d_0$.

\textbf{Step 3.}  Set the correction coefficients and form
$p_{\rm corr}$:
\begin{align*}
  c_j^{\rm corr}
    &= \sum_{k=1}^K \alpha_k\,\lambda_k\,T_{2j+1}(\lambda_k),\\
    &\qquad j = 0,\ldots,n_0-1,\\
  p_{\rm corr}(x)
    &= \sum_{j=0}^{n_0-1} c_j^{\rm corr}\, T_{2j+1}(x).
\end{align*}

\textbf{Step 4.}  Form the corrected polynomial:
\begin{align*}
  p_{SC}(x) = p_0(x) + p_{\rm corr}(x),
\end{align*}
where $p_{SC}$ has the \textbf{same degree} as $p_0$ and satisfies
$\lambda_k\, p_{SC}(\lambda_k) = 1$ exactly for $k = 1,\ldots,K$.
\end{keybox}

\begin{proposition}
\label{prop:bound}
Let $p_0$ satisfy $\abs{x\,p_0(x)-1} \leq \varepsilon$ on
$[a,1]$, and let $p_{SC} = p_0 + p_{\rm corr}$ be the corrected
polynomial. Let
$\Lambda_K = \mathrm{diag}(\lambda_1,\ldots,\lambda_K)$ and
$\mathbf{B}_K \in \R^{K \times n_0}$ with
$(B_K)_{kj} = T_{2j+1}(\lambda_k)$. If
$\lambda_k \in [a,1]$ for all $k$, then
\begin{equation}
  \abs{x\,p_{SC}(x) - 1} \leq \varepsilon
    + \norm{\Lambda_K\,\mathbf{B}_K}
    \cdot \norm{\boldsymbol\alpha}_2
    \cdot \norm{\mathbf{T}(x)}_2
\end{equation}
where
$\mathbf{T}(x) = (T_1(x), T_3(x), \ldots, T_{2n_0-1}(x))^T$.
\end{proposition}

\subsection{Degenerate eigenvalues}
\label{sec:degeneracy}

When the target set contains repeated or nearly repeated
eigenvalues --- as occurs naturally in problems with spatial
symmetry, such as the 2D Poisson equation on a square domain
where $\lambda_{j,k} = \lambda_{k,j}$ --- the corresponding rows
of $\Lambda_K \mathbf{B}_K$ are identical, rendering the Gram
matrix $\mathbf{G}$ rank-deficient. The truncated SVD solve
handles this algebraically, but the redundant constraints waste
degrees of freedom in the correction, producing a larger
$p_{\rm corr}$ than necessary and potentially degrading
uncorrected eigenvalues.

The remedy is straightforward: before applying the correction,
merge eigenvalues that are closer than a tolerance
$\delta_{\rm merge}$:
\begin{equation}
  |\lambda_i - \lambda_j| < \delta_{\rm merge}
  \;\Longrightarrow\;
  \text{retain one representative.}
  \label{eq:merge}
\end{equation}
Since a single interpolation constraint
$\lambda_k p_{SC}(\lambda_k) = 1$ applies equally to all eigenmodes
sharing the same eigenvalue, duplication removal loses no information.
The effective number of constraints $K_{\rm eff} \leq K$ equals
the number of distinct eigenvalues in the target set.

\subsection{Illustrative examples}
\label{sec:spectral_example}

We consider three examples to illustrate the spectrally
spectral correction.

\paragraph{Example 1:}
For $\kappa = 10$, $\varepsilon = 0.2$, we first construct the three base polynomials: Mang, Remez,
and S\"{u}nderhauf. We then let $\lambda_1 = 0.1$, $\lambda_2 = 0.5$, $\lambda_3 = 1.0$, i.e., $K = 3$, and construct the three corresponding corrected polynomials. In each case, $p_{SC}$ has
the \emph{same degree} as the base polynomial $p_0$, and the correction
requires solving only a $3 \times 3$ linear system~\eqref{eq:gram}.

\begin{figure*}[htbp]
  \centering
   \subcaptionbox{Remez and Spectral-Remez.}
  {\includegraphics[width=0.29\linewidth]{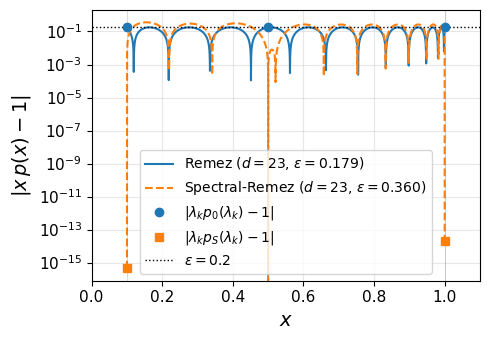}}
  \hfill
  \subcaptionbox{Mang and Spectral-Mang.}
  {\includegraphics[width=0.29\linewidth]{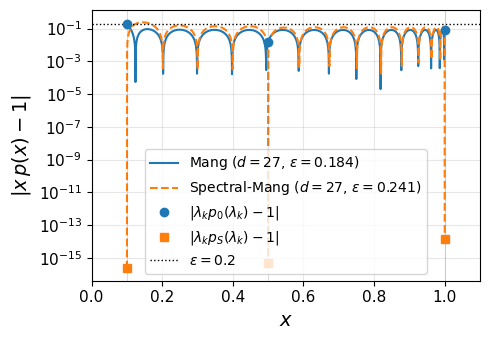}}
  \hfill
  \subcaptionbox{S\"{u}nderhauf and Spectral-S\"{u}nderhauf.}
  {\includegraphics[width=0.29\linewidth]{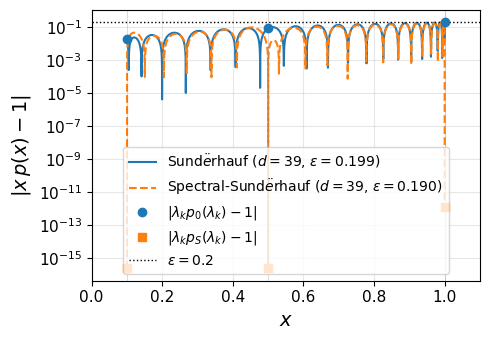}}
  \caption{Spectral correction applied to three base polynomials
  ($\kappa=10$, $\varepsilon=0.2$, $K=3$ eigenvalues at
  $\lambda = 0.1, 0.5, 1.0$). In all cases the spectrally corrected
  polynomial $p_{SC}$ achieves machine-precision errors at
  the eigenvalues, while
  preserving the continuous error profile of the base polynomial, at no increase in degree.}
  \label{fig:spectral_example1}
\end{figure*}

Figure~\ref{fig:spectral_example1} shows the point-wise error $|xp(x)-1|$
before and after correction for all three base polynomials. In every
case the spectral correction reduces the error at the three eigenvalues
to machine precision. Between eigenvalues, the spectral error closely tracks
the base but increases slightly as expected (consistent with Proposition \ref{prop:bound}). Notably, the spectral-Remez continuous error ($\varepsilon = 0.36$) exceeds that of the base Remez ($\varepsilon = 0.18$), confirming that the equioscillatory profile is most sensitive to perturbation. The spectral-S\"{u}nderhauf continuous error ($\varepsilon = 0.19$) remains essentially unchanged, reflecting the margin available in
the higher-degree base. However, recall that errors between eigenvalues do not affect the QSVT output, since the output state  $p(\bA)\bb = \sum_k p(\lambda_k)(\bv_k^T\bb)\bv_k$ depends only on the error of
$p$ at the eigenvalues of $\bA$.

\begin{figure*}[htbp]
  \centering
   \subcaptionbox{Remez and Spectral-Remez.}
  {\includegraphics[width=0.29\linewidth]{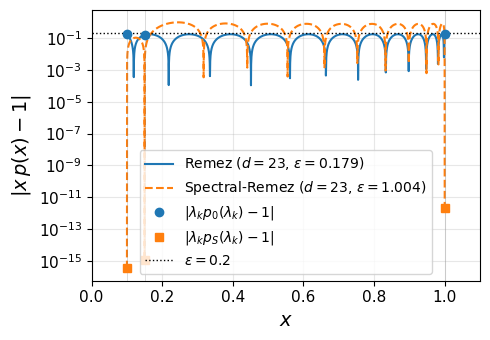}}
  \hfill
  \subcaptionbox{Mang and Spectral-Mang.}
  {\includegraphics[width=0.29\linewidth]{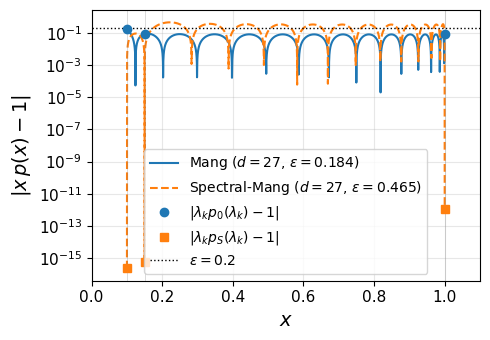}}
  \hfill
  \subcaptionbox{S\"{u}nderhauf and Spectral-S\"{u}nderhauf.}
  {\includegraphics[width=0.29\linewidth]{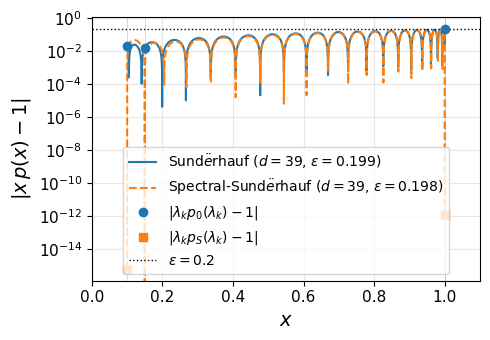}}
  \caption{Spectral correction applied to three base polynomials
  ($\kappa=10$, $\varepsilon=0.2$, $K=3$ eigenvalues at
  $\lambda = 0.1, 0.15, 1.0$). The spectrally corrected polynomial achieves machine-precision errors at the eigenvalues, but the closely spaced eigenvalues cause a perceptible increase in the inter-eigenvalue
error. }
  \label{fig:spectral_example2}
\end{figure*}

\paragraph{Example 2: }
With the same base polynomials as the previous example, we set the second eigenvalue closer to the lowest: $\lambda_1 = 0.1$, $\lambda_2 = 0.15$, $\lambda_3 = 1.0$. Figure~\ref{fig:spectral_example2} shows the point-wise error $|xp(x)-1|$ before and after correction for all three base polynomials.  While the closely spaced pair $\lambda_1 = 0.1$,
$\lambda_2 = 0.15$ causes a significant increase in the
inter-eigenvalue error, this will not affect the QSVT output,
which depends on $p$ only at the eigenvalues. Furthermore, the subnormalization factor increases only modestly from $\tau_0 \in [8.2, 9.8]$ to $\tau_{SC} = \kappa = 10.0$ for all three bases, since the correction enforces $p_{SC}(\lambda_1) = 1/\lambda_1$ exactly. The success probability $1/\tau^2$ is therefore governed by $\kappa$ alone, independent of the base polynomial.

\begin{figure*}[htbp]
  \centering
   \subcaptionbox{Remez and spectral-Remez.}
  {\includegraphics[width=0.29\linewidth]{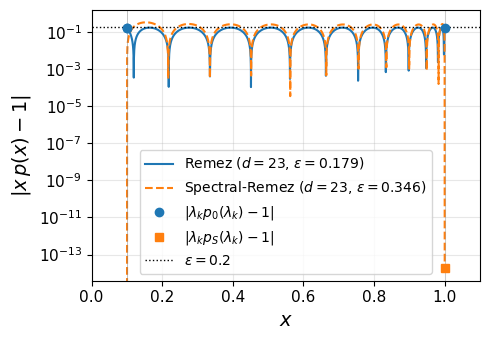}}
  \hfill
  \subcaptionbox{Mang and spectral-Mang.}
  {\includegraphics[width=0.29\linewidth]{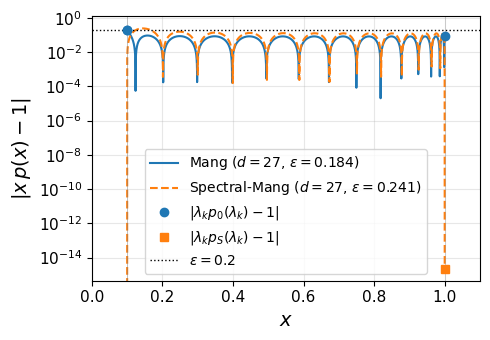}}
  \hfill
  \subcaptionbox{S\"{u}nderhauf and spectral-S\"{u}nderhauf.}
  {\includegraphics[width=0.29\linewidth]{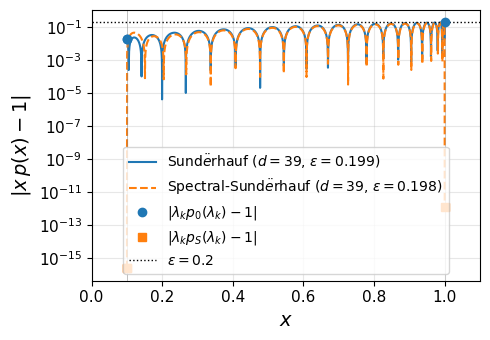}}
  \caption{Spectral correction with a degenerate eigenvalue pair
  ($\kappa=10$, $\varepsilon=0.2$, $\lambda = 0.1, 0.1, 1.0$,
  $K_{\rm eff} = 2$ after duplication removal). The correction achieves
  machine-precision errors at both distinct eigenvalues, with the
  continuous error profile matching Example~1.}
  \label{fig:spectral_example3}
\end{figure*}

\paragraph{Example 3: }
Next, we set the second eigenvalue identical to the first:
$\lambda_1 = 0.1$, $\lambda_2 = 0.1$, $\lambda_3 = 1.0$, to
illustrate the effect of degenerate eigenvalues. The duplication removal procedure
of Section~\ref{sec:degeneracy} merges the repeated pair,
reducing the effective correction to $K_{\rm eff} = 2$ distinct
eigenvalues. Figure~\ref{fig:spectral_example3} shows the
point-wise error $|xp(x)-1|$. The results are visually identical
to Example~1 ($\lambda = 0.1, 0.5, 1.0$ with $K=3$), confirming
that duplication removal correctly handles repeated eigenvalues without
degradation.

\section{Numerical Experiments}
\label{sec:experiments}
Having established the polynomial-level behavior of the spectral correction, we now validate the proposed formulations on 1D and 2D Poisson problems. 

\subsection {1D and 2D Poisson Finite Difference}
For convenience, we provide here the well-known analytical expressions for the eigenvalues of finite difference formulations of 1D and 2D Poisson problems.

For the 1D Poisson equation $-u'' = f$ on $(0,1)$ with Dirichlet
boundary conditions, discretized on $N$ interior nodes with
$h = 1/(N+1)$, the eigenvalues are~\cite{golub2013matrix}
\begin{align}
  \lambda_k &= \frac{4}{h^2}\sin^2\!\left(\frac{k\pi}{2(N+1)}\right),\nonumber\\
  &\qquad k = 1, \ldots, N,
  \label{eq:eig_fd_1d}
\end{align}
with condition number $\kappa \approx \frac{4}{\pi^2}(N+1)^2$.

For the 2D Poisson equation $-\nabla^2 u = f$ on $(0,1)^2$
with Dirichlet boundary conditions, discretized with $N_1$
interior nodes per direction ($N = N_1^2$), the eigenvalues are pairwise sums of the 1D
eigenvalues~\cite{golub2013matrix}:
\begin{align}
  \lambda_{j,k} &= \frac{4}{h^2}\biggl[
    \sin^2\!\biggl(\frac{j\pi}{2(N_1\!+\!1)}\biggr)
    + \sin^2\!\biggl(\frac{k\pi}{2(N_1\!+\!1)}\biggr)
  \biggr], \nonumber\\
  &\qquad j,k = 1, \ldots, N_1,
  \label{eq:eig_fd_2d}
\end{align}
with the same condition number $\kappa \approx
\frac{4}{\pi^2}(N_1+1)^2$ as the 1D case. 

In both cases, the normalized eigenvalues $\tilde\lambda = \lambda/\lambda_{\max}$
satisfy $\tilde\lambda \in (0,1]$ with
$\tilde\lambda_{\min} = 1/\kappa$.

\subsection{QSVT Metrics}

Given the block-encoded matrix $\bA$ and a polynomial approximation to $1/x$, the QSP phase angles $\{\phi_j\}_{j=0}^d$ are computed via the \texttt{pyqsp} package~\cite{haah2019product, gilyen2019quantum}, and the resulting quantum circuit is simulated via \emph{noiseless statevector}
simulation in Qiskit~\cite{javadi2024quantum}. 

If $\bu = \bA^{-1}\bb / \|\bA^{-1}\bb\|$ denotes the normalized classical solution and $\bu_{\rm QSVT}$ the normalized QSVT output,
the output state fidelity is
\begin{equation}
  F = \left|\bu^\top \bu_{\rm QSVT}\right|^2,
  \label{eq:fidelity}
\end{equation}
The classical compliance $C = \bb^\top \bA^{-1}\bb$ and the QSVT compliance is recovered as
\begin{equation}
  C_{\rm QSVT} = (\bb^\top \bu_{\rm QSVT})\,\tau\,\sqrt{P_{\rm succ}},
  \label{eq:compliance_qsvt}
\end{equation}
where $\tau$ is the polynomial subnormalization factor~\eqref{eq:tauDefinition},
and $P_{\rm succ}$ is the post-selection success
probability~\eqref{eq:psucc}. The relative compliance error $|C_{\rm QSVT} - C|/|C|$ provides a
physically meaningful scalar measure of solution accuracy, in addition to the state fidelity~$F$.

\subsection{Pure Spectral Polynomial}
\label{sec:exp_spectral}

We first assess the pure spectral polynomial $p_S$ of
Section~\ref{sec:spectralpoly}, which requires all $N$
eigenvalues and is applicable when $N$ is small. 
Table~\ref{tab:pure_spectral_1d} reports the degree--$\tau$
trade-off for the 1D Poisson problem with $m=3$ ($N=8$,
$\kappa = 32.5$) and $m=4$ ($N=16$, $\kappa = 117.6$).

\begin{table}[htbp]
\centering
\caption{Pure spectral polynomial for the 1D Poisson equation.
All cases achieve unit fidelity ($F = 1.000000$) with uniform
load; only $\tau/\kappa$ and the success probability vary.}
\label{tab:pure_spectral_1d}
\resizebox{\columnwidth}{!}{%
\begin{tabular}{rrrrrrr}
\toprule
& \multicolumn{3}{c}{$m=3$ ($N=8$, $\kappa=32.5$)}
& \multicolumn{3}{c}{$m=4$ ($N=16$, $\kappa=117.6$)} \\
\cmidrule(lr){2-4} \cmidrule(lr){5-7}
$n_{\rm factor}$ & $d$ & $\tau/\kappa$ & $P_{\rm succ}$
& $d$ & $\tau/\kappa$ & $P_{\rm succ}$ \\
\midrule
2   & 31  & 1.41 & 0.577  & 63  & 28.5 & 0.324 \\
3   & 47  & 1.25 & 0.728  & 95  & 1.63 & 0.500 \\
4   & 63  & 1.01 & 0.929  & 127 & 1.39 & 0.533 \\
5   & 79  & 1.01 & 0.967  & 159 & 1.30 & 0.663 \\
8   & 127 & 1.00 & 0.979  & 255 & 1.00 & 0.887 \\
\bottomrule
\end{tabular}%
}
\end{table}

For $m=3$, $n_{\rm factor} = 4$ ($d=63$) suffices to reach
$\tau \approx \kappa$. For $m=4$, $n_{\rm factor} = 8$ ($d=255$)
is required --- comparable to the base Mang degree at
$\varepsilon = 0.2$ ($d = 305$). This confirms the second
limitation of Section~\ref{sec:spectralpoly}: the $n_{\rm factor}$
needed for $\tau \approx \kappa$ grows with $\kappa$, eventually
eroding the degree advantage over base polynomials.

The pure spectral polynomial is therefore most effective for
small problems where all eigenvalues are known and $\kappa$ is
moderate. For larger or higher-$\kappa$ problems, the spectrally
corrected approach of the following subsections is preferred.

\subsection{Spectrally Corrected Polynomials}

\subsubsection{1D Poisson Polynomial Error}

\begin{figure*}[htbp]
  \centering
   \subcaptionbox{Remez and Spectral-Remez.}
  {\includegraphics[width=0.29\linewidth]{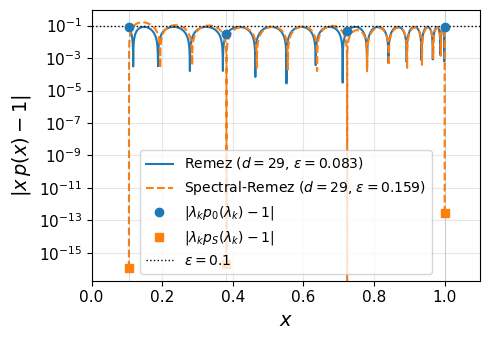}}
  \hfill
  \subcaptionbox{Mang and Spectral-Mang.}
  {\includegraphics[width=0.29\linewidth]{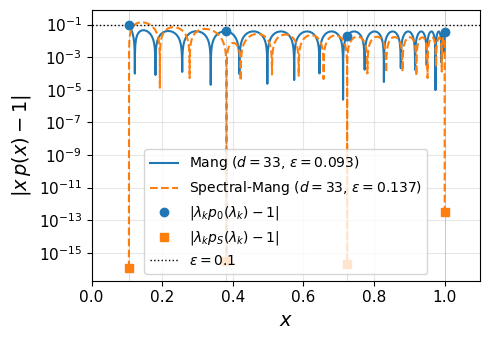}}
  \hfill
  \subcaptionbox{S\"{u}nderhauf and Spectral-S\"{u}nderhauf.}
  {\includegraphics[width=0.29\linewidth]{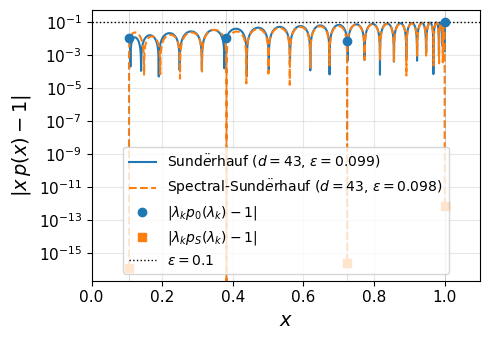}}
  \caption{Spectral correction using all $K=4$ eigenvalues of the 1D
Poisson FD matrix with $N=4$ interior nodes ($\kappa=9.47$,
$\varepsilon=0.1$, $\lambda_1=0.106$, $\lambda_2=0.382$,
$\lambda_3=0.724$, $\lambda_4=1.0$). }
  \label{fig:spectral_1d_poisson}
\end{figure*}

We set $K = N = 4$, using all eigenvalues of the 1D Poisson FD matrix
with $N=4$ interior nodes ($\kappa = 9.47$, $\varepsilon = 0.1$):
$\lambda_1 = 0.1056$, $\lambda_2 = 0.3820$, $\lambda_3 = 0.7236$,
$\lambda_4 = 1.0$. Figure~\ref{fig:spectral_1d_poisson} shows that spectral correction for all three base polynomials achieve machine-precision errors at all four eigenvalues, with the
spectral error closely tracking the base between eigenvalues.
This confirms that when the full eigenvalue spectrum is available and
well-separated, the spectral correction is effective regardless of the
choice of base polynomial.

\subsubsection{Degree--accuracy trade-off}
\label{sec:exp_degree}

A central benefit of the spectral correction is that it
achieves machine-precision eigenvalue accuracy at a much lower degree
than a base polynomial. Eigenvalue accuracy is measured via the maximum eigenvalue residual:
\begin{align}
    \mathcal{E}_{\rm eig} = \max_k |\lambda_k p(\lambda_k) - 1|.
\end{align}
For brevity, we use Mang as a base polynomial in this experiment; results for Remez and S\"{u}nderhauf are qualitatively similar. Table~\ref{tab:degree_accuracy} makes the central claim explicit for $N = 4$
($\kap \approx 9.47$) with $K = 2 = N/2$ corrected eigenvalues.

\begin{table}[htbp]
\centering
\caption{Degree $d$ and maximum eigenvalue residual
$\mathcal{E}_{\rm eig} = \max_k |\lambda_k p(\lambda_k) - 1|$
for Mang and spectral-Mang at increasing accuracy targets.
$N=4$, $\kap \approx 9.47$, $K=2$. Two columns report
$\mathcal{E}_{\rm eig}$ at the $K$ corrected eigenvalues and
over the full spectrum respectively.}
\label{tab:degree_accuracy}
\resizebox{\columnwidth}{!}{%
\begin{tabular}{llrrr}
\toprule
Method & $\varepsilon$ & $d$
  & $\mathcal{E}_{\rm eig}$ ($K$ corr.)
  & $\mathcal{E}_{\rm eig}$ (all $N$) \\
\midrule
Mang                    & $0.2$    & $25$ & $1.92\times10^{-1}$ & $1.92\times10^{-1}$ \\
Mang                    & $0.1$    & $33$ & $9.27\times10^{-2}$ & $9.27\times10^{-2}$ \\
Mang                    & $0.01$   & $57$ & $9.05\times10^{-3}$ & $9.05\times10^{-3}$ \\
\midrule
Spectral-Mang ($K=2$)   & $0.2$    & $25$ & $2.22\times10^{-16}$ & $1.58\times10^{-1}$ \\
Spectral-Mang ($K=2$)   & $0.1$    & $33$ & $1.11\times10^{-16}$ & $1.51\times10^{-2}$ \\
Spectral-Mang ($K=2$)   & $0.01$   & $57$ & $1.00\times10^{-16}$ & $2.71\times10^{-3}$ \\
\bottomrule
\end{tabular}%
}
\end{table}

Two observations follow from Table~\ref{tab:degree_accuracy}.
First, spectral-Mang with $K=2$ achieves machine-precision
residuals at the corrected eigenvalues, regardless of
$\varepsilon$, while the base polynomial requires
$\varepsilon \to 0$ (and correspondingly higher degree) to
approach the same accuracy. Second, the uncorrected eigenvalues retain the accuracy of
the base polynomial, consistent with
Proposition~\ref{prop:bound}.

\subsubsection{QSVT validation for 1D Poisson}
\label{sec:exp_qsvt}

We now validate the full QSVT pipeline using the spectrally
corrected polynomial. 

Table~\ref{tab:qsvt_results} reports results for $m=4$ ($N=16$,
$\kappa=117.6$) with $K=N=16$ corrected eigenvalues, using the
Mang base polynomial. Two load cases are shown: a uniform load
(smooth, low-frequency dominated) and a point load at the midpoint
(broadband, all modes excited). 
\begin{table}[htbp]
\centering
\caption{QSVT results for the 1D Poisson equation, $m=4$ ($N=16$,
$\kappa=117.6$), $K=N=16$, Mang base. The spectrally corrected
method at $\varepsilon=0.5$ achieves unit fidelity and lower
compliance error than the tight reference at $5.28\times$ lower
circuit depth.}
\label{tab:qsvt_results}
\resizebox{\columnwidth}{!}{%
\begin{tabular}{lrrrrr}
\toprule
Method & $d$ & Fidelity & Rel.\ compl.\ err
  & $P_{\rm succ}$ & $\tau$ \\
\midrule
\multicolumn{6}{l}{\textit{Uniform load}} \\
\midrule
Mang ($\varepsilon=0.5$)
  & 177 & 0.999536 & $4.91\times10^{-1}$ & 0.772 & 74.4 \\
Mang ($\varepsilon=10^{-3}$)
  & 935 & 1.000000 & $5.78\times10^{-4}$ & 0.717 & 156.2 \\
Spectral-Mang ($\varepsilon=0.5$)
  & 177 & 1.000000 & $3.73\times10^{-5}$ & 0.781 & 142.8 \\
\midrule
\multicolumn{6}{l}{\textit{Point load at midpoint}} \\
\midrule
Mang ($\varepsilon=0.5$)
  & 177 & 0.991581 & $4.09\times10^{-1}$ & 0.667 & 74.4 \\
Mang ($\varepsilon=10^{-3}$)
  & 935 & 1.000000 & $5.25\times10^{-4}$ & 0.627 & 156.2 \\
Spectral-Mang ($\varepsilon=0.5$)
  & 177 & 1.000000 & $1.04\times10^{-4}$ & 0.664 & 142.8 \\
\bottomrule
\end{tabular}%
}
\end{table}

The spectral correction achieves unit fidelity for both load cases
at $\varepsilon = 0.5$, matching the tight reference
($\varepsilon = 10^{-3}$) at $5.28\times$ lower circuit depth
($d=177$ vs $d=935$). The compliance error is an order of magnitude
better than the tight reference for the uniform load
($3.73\times10^{-5}$ vs $5.78\times10^{-4}$) and comparable for
the point load ($1.04\times10^{-4}$ vs $5.25\times10^{-4}$). 
The success probability of the spectrally corrected methods
is comparable to the loose base despite a larger polynomial
subnormalization factor $\tau$, because the improved polynomial
accuracy increases the numerator $\|p_{SC}(\bA)\bb\|^2$
proportionally.  Spectral-Remez and spectral-S\"{u}nderhauf yield qualitatively identical results and are not included here for brevity.

\subsubsection{Robustness to eigenvalue perturbations}
\label{sec:exp_perturbation}

In practice, eigenvalues may be known only approximately, for
example via a Lanczos iteration rather than the closed-form
expression~\eqref{eq:eig_fd_1d}. We assess robustness by
replacing the exact eigenvalues $\lambda_k$ with perturbed values
$\hat\lambda_k = \lambda_k(1 + \delta_k)$,
$\delta_k \sim \mathcal{U}(-\eta, \eta)$, and repeating the QSVT
experiment of Section~\ref{sec:exp_qsvt}, for the uniform load case, with $n=10$ independent
trials per perturbation level. 
Table~\ref{tab:perturbation} reports the mean and standard
deviation of fidelity and relative compliance error.

\begin{table}[htbp]
\centering
\caption{Robustness of spectral-Mang to eigenvalue perturbations.
$m=4$ ($N=16$, $\kappa=117.6$), $K=N=16$, $\varepsilon=0.5$,
$n=10$ trials per perturbation level $\eta$.}
\label{tab:perturbation}
\resizebox{\columnwidth}{!}{%
\begin{tabular}{lrrr}
\toprule
$\eta$ & Fidelity & Rel.\ compl.\ err & $P_{\rm succ}$ \\
\midrule
$0$ (exact)  & $1.000000$
  & $1.56\times10^{-4}$
  & $0.821$ \\
$10^{-2}$    & $1.000000 \pm 3\times10^{-7}$
  & $3.32\times10^{-3} \pm 5\times10^{-3}$
  & $0.821$ \\
$10^{-1}$    & $0.999808 \pm 3\times10^{-4}$
  & $1.81\times10^{-2} \pm 4\times10^{-2}$
  & $0.822$ \\
\bottomrule
\end{tabular}%
}
\end{table}

Fidelity remains at unity to five decimal places at
$\eta = 10^{-2}$, and even at $\eta = 10^{-1}$, the fidelity loss is below
$2\times10^{-4}$ and the compliance error remains below
$2\times10^{-2}$. The success probability is unaffected by
eigenvalue perturbation, confirming that $\tau$ depends only on
the base polynomial.

\subsubsection{QSVT validation for 2D Poisson }
\label{sec:exp_2d}
A uniform load is used as the forcing function for the 2D Poisson problem and a full QSVT pipeline is executed. Table~\ref{tab:qsvt_2d}
reports fidelity and peak solution value for spectral-Mang at
$\varepsilon = 0.2$ ($d = 305$) with increasing $K$. The base Mang polynomial already achieves high fidelity ($0.99987$) for this
smooth load; the spectral correction pushes fidelity to
$0.9999996$ at $K=32$ ($K_{\rm eff}=18$ after removing duplicates),
recovering the classical peak value to within $0.02\%$.

Unlike the 1D case where $K = N$ corrects the full spectrum, the
2D problem has $N = N_1^2$ eigenvalues but the polynomial has only
$n_0 = (d+1)/2 = 153$ odd Chebyshev terms, limiting the effective
number of corrections to $K \ll n_0$. In practice, correcting the
$K \approx 8$--$16$ smallest eigenvalues --- where the base
polynomial error is largest --- suffices to achieve fidelity above
$0.9999$.

\begin{table}[htbp]
\centering
\caption{Spectral-Mang on the 2D Poisson equation ($N=256$,
$\kappa=117.6$, $\varepsilon=0.2$, $d=305$, uniform load).
Classical peak value: $0.1044$.}
\label{tab:qsvt_2d}
\resizebox{\columnwidth}{!}{%
\begin{tabular}{rrrrr}
\toprule
$K$ & $K_{\rm eff}$ & Fidelity & Max value & Peak error \\
\midrule
0 (base) & --  & 0.99987 & 0.1033 & $-1.1\%$ \\
1        & 1   & 0.99969 & 0.1071 & $+2.5\%$ \\
4        & 3   & 0.99941 & 0.1081 & $+3.5\%$ \\
8        & 5   & 0.99997 & 0.1040 & $-0.4\%$ \\
16       & 10  & 0.999998 & 0.1046 & $+0.2\%$ \\
32       & 18  & 0.9999996 & 0.1044 & $+0.0\%$ \\
\bottomrule
\end{tabular}%
}
\end{table}

\begin{figure*}[htbp]
  \centering
  \subcaptionbox{Classical; max $= 0.1044$.}
    {\includegraphics[width=0.30\linewidth]{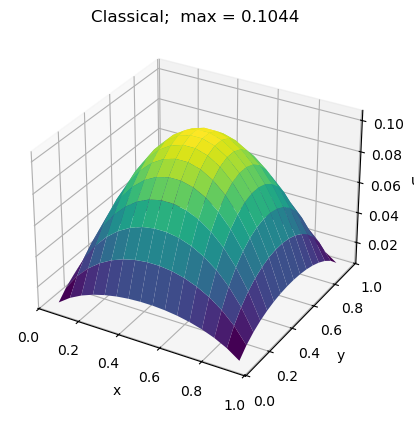}}
  \hfill
  \subcaptionbox{Mang; max $= 0.1033$.}
    {\includegraphics[width=0.30\linewidth]{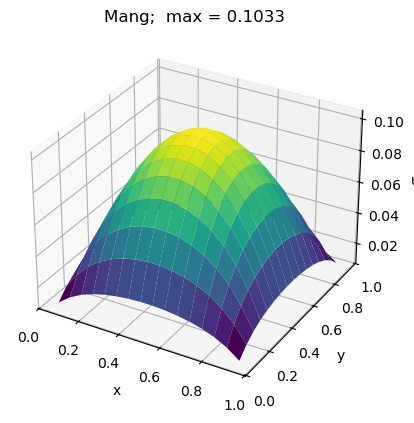}}
  \hfill
  \subcaptionbox{Spectral-Mang; max $= 0.1044$.}
    {\includegraphics[width=0.30\linewidth]{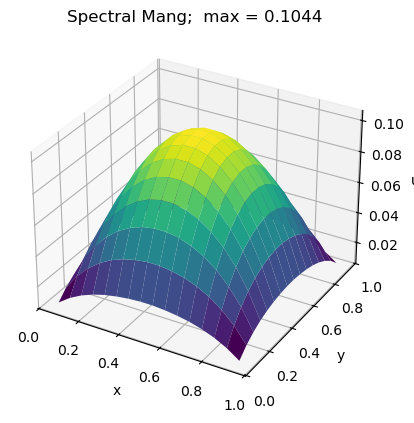}}
  \caption{2D Poisson solution surfaces ($N=256$, $\kappa=117.6$,
  uniform load, $\varepsilon=0.2$, $d=305$). Spectral-Mang with
  $K=32$ ($K_{\rm eff}=18$) recovers the classical solution.}
  \label{fig:2d_surfaces}
\end{figure*}

Figure~\ref{fig:2d_surfaces} shows the solution surfaces for the
classical, base Mang, and spectral-Mang ($K=32$) cases.

\section{Discussion}
\label{sec:discussion}

\subsection{Scope and limitations}
\label{sec:limitations}

The approach is most effective for structured discretizations where
eigenvalues are available analytically, such as FD and FEM on uniform grids.
 For unstructured meshes, a preprocessing step (Lanczos or QPE) is required,
and its cost must be weighed against the QSVT benefit.  

Furthermore, when eigenvalues are obtained approximately, the correction enforces
$\hat\lambda_k p_{SC}(\hat\lambda_k) = 1$ exactly at the \emph{supplied}
values, not at the true eigenvalues of $\bA$. The accuracy of the
QSVT output, therefore, depends on how closely $\hat\lambda_k$
approximates $\lambda_k$. The perturbation experiment of
Section~\ref{sec:exp_perturbation} quantifies this dependence:
errors up to $\eta = 10\%$ degrade the compliance error
proportionally while fidelity loss remains below $2\times10^{-5}$.

\subsection{Relation to classical polynomial preconditioning}
\label{sec:classical}
Exploiting eigenvalue estimates to accelerate linear solvers has a long history in classical polynomial preconditioning~\cite{fischer1994adaptive, ashby1992comparison,
saad1985practical, loe2022polynomial, zhang2025efficient}.  The spectral correction presented here is the natural QSVT analogue: eigenvalue knowledge is used to shape the polynomial, thereby improving accuracy at a fixed circuit depth. The main difference is that, in a classical setting, polynomial preconditioning has largely been replaced by more efficient preconditioners such as ILU. On the other hand, in the QSVT setting, polynomial approximation is mandatory, making spectral adaptation both natural and necessary. Furthermore, classical polynomial preconditioning relies on eigenvalue estimates; we have exploited exact, closed-form expressions to establish the method and its limits.  When closed-form eigenvalues are unavailable, the perturbation experiment of Section~\ref{sec:exp_perturbation} suggests that approximate eigenvalues may suffice.

\subsection{Future directions}
\label{sec:future}

\paragraph{Parameter sensitivity:}
The interplay between the base tolerance $\varepsilon$, the number
of corrected eigenvalues $K$, the polynomial degree $d$, and the eigenvalue density of the operator warrants further investigation.
The 2D experiments in Section~\ref{sec:exp_2d} suggest that denser
spectra require tighter base tolerances to avoid degrading
uncorrected eigenvalues, but a systematic characterization of this
trade-off remains open.

\paragraph{Extension to FEM and elasticity operators:}
The experiments in this paper use finite difference discretizations
of the Poisson equation. Extending the spectral correction to
finite element discretizations and to the Lam\'{e} equations of linear elasticity is a natural next step. For operators with tensor-product structure, closed-form eigenvalues are available.

\paragraph{Lanczos and QPE eigenvalue extraction:}
For operators without closed-form eigenvalues, a short Lanczos run
($\sim 20$ steps) provides the $K$ smallest Ritz values to $\sim 0.1\%$
accuracy~\cite{golub2013matrix}, sufficient for the spectral correction given its demonstrated robustness to eigenvalue perturbations. Alternatively, quantum phase estimation can
extract eigenvalues from the block encoding directly, amortizing the
cost over multiple solves and eliminating the classical preprocessing
step entirely.

\paragraph{Spectral extension:}
An alternative to the spectral correction is \emph{spectral extension} where we extend the base
polynomial by $K$ new odd Chebyshev terms of degree
$d_0+2, d_0+4, \ldots, d_0+2K$, using those additional degrees of
freedom exclusively to satisfy the $K$ interpolation constraints. This preserves the continuous error profile of the base polynomial exactly --- including the equioscillation of the Remez polynomial ---
at the cost of increasing the degree by $2K$.  The square $K\times K$ system is solved for the new
coefficients alone, leaving the existing coefficients untouched. Spectral extension and
spectral correction fall under the umbrella of \emph{spectral bootstrapping}: exploiting
eigenvalue knowledge to improve QSVT polynomials.

\paragraph{QSVT on near-term hardware:}
The circuit depth reductions demonstrated here (up to $5.28\times$
for the 1D Poisson problem) directly reduce error accumulation on
near-term hardware. Combining the spectral correction with
zero-noise extrapolation~\cite{temme2017error} could further close the gap between statevector
fidelity and hardware fidelity, making QSVT of small Poisson instances
($N=4$, $d=25$) tractable on current quantum devices.

\section{Conclusion}
\label{sec:conclusion}

We have presented a framework for reducing polynomial degree ---
and hence quantum circuit depth --- in QSVT-based linear solvers
by exploiting prior knowledge of the matrix spectrum.

The key observation is that QSVT solution accuracy depends only on
the polynomial values at the eigenvalues of $\bA$, not between
them. When all $N$ eigenvalues are known, a pure spectral
polynomial achieves unit fidelity at reduced degree, but has practical limitations.

To address these limitations, we propose the \emph{spectral correction}: given any base polynomial $p_0$ of
degree $d_0$ and $K \leq N$ known eigenvalues, a $K \times K$
Gram system enforces exact interpolation at those eigenvalues
without increasing $d_0$. The corrected polynomial $p_{SC}$
achieves machine-precision residuals at the corrected eigenvalues
regardless of the base tolerance $\varepsilon$, while the base
polynomial provides a continuous safety net at uncorrected
eigenvalues.

QSVT experiments on the 1D Poisson equation ($N=16$,
$\kappa=117.6$) demonstrate up to a $5.28\times$ reduction in
circuit depth relative to a base Mang polynomial, at unit
fidelity and improved compliance error. Spectral correction is agnostic
to the choice of base polynomial and robust to eigenvalue
perturbations up to $10\%$ relative error. Extension to the 2D
Poisson equation ($N=256$) confirms that correcting a small
fraction of the spectrum ($K_{\rm eff} = 10$ of $256$ eigenvalues)
suffices to achieve fidelity above $0.999$, with degenerate
eigenvalues handled via duplication removal. Future directions include
extension to FEM and elasticity operators, systematic
characterization of the tolerance--density trade-off, and
Lanczos/QPE eigenvalue extraction as outlined in
Section~\ref{sec:future}.

\section*{Acknowledgments}
The author would like to acknowledge the Vilas Associate Grant from the University of Wisconsin Graduate School.

\section*{Declarations}
The author declares no conflict of interest.

\section*{Code reproducibility and availability }

The code developed in this work is available at
\url{https://github.com/UW-ERSL/SpectralCorrectionQSVT.git}
and will be made public upon publication.

\bibliographystyle{unsrtnat}
\bibliography{refs}

\end{document}